\documentstyle[epsfig,12pt]{article}
\newcommand{\bce}{\begin{center}}
\newcommand{\ece}{\end{center}}
\newcommand{\beq}{\begin{equation}}
\newcommand{\eeq}{\end{equation}}
\newcommand{\bea}{\vspace{0.25cm}\begin{eqnarray}}
\newcommand{\eea}{\end{eqnarray}}

\newcommand{\bsigma}{\mbox{\boldmath $\sigma$}}

\newcommand{\ba}{\begin{array}}
\newcommand{\ea}{\end{array}}

\def\lsim{\mathrel{\rlap{\lower4pt\hbox{\hskip1pt$\sim$}}
    \raise1pt\hbox{$<$}}}         
\def\gsim{\mathrel{\rlap{\lower4pt\hbox{\hskip1pt$\sim$}}
    \raise1pt\hbox{$>$}}}         

\def\lsim{\mathrel{\rlap{\lower4pt\hbox{\hskip1pt$\sim$}}
    \raise1pt\hbox{$<$}}}         
\def\gsim{\mathrel{\rlap{\lower4pt\hbox{\hskip1pt$\sim$}}
    \raise1pt\hbox{$>$}}}         

  \advance\oddsidemargin  by -1in
  \advance\evensidemargin by -1in
  \advance\topmargin      by -0.5in
\def\lsim{\mathrel{\rlap{\lower4pt\hbox{\hskip1pt$\sim$}}
    \raise1pt\hbox{$<$}}}         
\def\gsim{\mathrel{\rlap{\lower4pt\hbox{\hskip1pt$\sim$}}
    \raise1pt\hbox{$>$}}}         

\def\beq{\begin{equation}}
\def\endeq{\end{equation}}
\def\arr{\begin{eqnarray}}
\def\endarr{\end{eqnarray}}
\makeindex
 
\begin{document}

\begin{flushright}
ITEP-PH-2/2004
\end{flushright}
\vspace{1cm}

\begin{center}
{\Large \bf
Ultra-Relativistic Nuclei in Crystal Channel:
Coulomb Scattering, Coherence and Absorption

 \vspace{1.0cm}}
 
{\large \bf
V.R. Zoller\medskip\\ }
{\sl  Institute for Theoretical and Experimental Physics \\
117218 Moscow Russia
{\footnote {\rm zoller@itep.ru}}\vspace{1cm}\\}
{\bf           Abstract}
\end{center}

We incorporate the effect of lattice thermal vibrations into 
 the Glauber theory description of 
 particle- and nucleus-crystal Coulomb interactions at high-energy.
The allowance for the lattice thermal vibrations  is shown to produce 
 strong  absorption effect: the  phase shift function of   
the multiple-diffraction  scattering on a chain of $N$ identical atoms 
acquires large imaginary part and the radius of the absorption region 
in the impact parameter plane grows logarithmically with $N$. 
Consequences of this
observation for the elastic and quasi-elastic  Coulomb scattering  are 
discussed. Practically interesting example of  the coherent Coulomb
excitation of ultra-relativistic particles and nuclei passing through the 
crystal is considered in detail.

\newpage

\section{Introduction}
 
In this paper we develop the Glauber theory \cite{GLAUBER} description 
 of the  absorption phenomenon in 
coherent  particle- and nucleus-crystal Coulomb interactions at high-energy.

As is well known, the  multi-loop corrections  generate the 
imaginary part of the scattering amplitude  even  if 
the tree-level amplitude is  purely real. For example, 
the purely real Born amplitude of the 
  high-energy Coulomb scattering in crystal
  acquires the imaginary part due to the multiple scattering (MS) effects.
 However, in the widely used
 static/frozen  lattice  
approximation (SL approximation)
 the account of rescatterings  alters only the overall real phase of 
the full amplitude  thus producing no absorption effect . 
The latter is related to the 
creation and annihilation of excited intermediate 
 states of crystal and as such
manifests itself only beyond the SL approximation
 (for the analysis of elastic scattering 
 based on the SL approximation see \cite{KALASH}). Indeed,
the amplitude of small-angle elastic scattering on a chain of $N$
identical atoms in the impact parameter representation  equals
$$1-\langle S(b)\rangle,$$
 where the scattering matrix placed between the 
ground states of crystal is 
$$\langle S(b)\rangle=\langle \exp[i\chi(b)]\rangle$$
 with the purely real
phase shift function $\chi(b)=\sum_{j=1}^N\chi_j(b)$.  In the SL approximation 
$$\langle \exp[i\chi(b)]\rangle
\simeq \exp[i\chi(b)].$$
Therefore, only  with   the allowance for the 
lattice thermal vibrations the Coulomb phase shift function gets, in general,
 non-vanishing imaginary part which is interpreted as an  absorption 
effect. The  imaginary part  appears only  as the second 
order perturbation, 
$$\sim {i\over 2}[\langle\chi^2\rangle-\langle\chi\rangle^2].$$
 But the strength of the effect is 
 $\propto\beta^2 N$, where 
$\beta$ is the coupling constant. 
For the coherent scattering of relativistic nuclei 
(the electric charge  $Z_1$)
on the chain of $N$ atoms (the atomic number  $Z_2$) in a crystal 
the effective coupling $\beta=2\alpha Z_1Z_2$ is strong and the absorption 
effect is  strong as well. The absorption is strong 
for  impact parameters, $b$, smaller than some characteristic value
 $b_a\propto\log(\beta N)$, and vanishes toward the region of larger $b$.
This phenomenon provides a natural ultra-violet regulator of the theory
and enables, in particular, consistent calculation
of  the coherent elastic scattering 
cross section. The latter is calculated  and turns out to be equal to the
one half of the total cross section. 
As we shall see, the  absorption effect is  of prime importance also
for   quantitative understanding of the phenomenon of the coherent Coulomb
excitation of relativistic particles and  nuclei passing through the crystal.
A consistent description of this phenomenon is the goal of our paper.

The outline of the paper is as follows. 
We start with 
the well known example of  the coherent Coulomb  elastic scattering 
of charged  particle/nucleus   by  a linear chain of $N$ identical atoms  
in a crystal target (Sec.2). 
We derive the scattering matrix with 
absorption and
calculate the cross section of the 
coherent elastic scattering, $\sigma_{el}$ (Sec. 3) and  the cross section
of the incoherent excitation and break-up of the target, $\sigma_{Qel}$ 
(Sec. 4).
We find that in the large-$N$ limit $\sigma_{el}\approx\sigma_{Qel}\approx
{1\over 2}\sigma_{tot}$ (Sec.5). In Sec.6 we discuss 
  the  coherent Coulomb excitation of ultra-relativistic
particles and  nuclei passing through the  crystal to the lowest order
of perturbation theory. The higher order effects we consider
in Sec. 7 where the cross section of the process is calculated.
We finally conclude with a brief summary in Sec. 8. 

\section{Coherent elastic scattering and absorption.}
 
The interatomic distances in crystal, $a$,  are large,  
compared to the  
 Thomas-Fermi screening radius $r_{0}$, 
$a\sim 3-5\AA\gg r_{0}= r_B Z_2^{-1/3}\sim 0.1\AA$,
 where $Z_2$ is the atomic number of the target atom and $r_B$
 is the Bohr radius \cite{GEM}. 
The relevant impact parameters, $b$, 
satisfy the condition  $b\ll a$ and  the amplitudes of 
scattering by different atomic chains  parallel to a given  crystallographic 
axis are incoherent.

The   amplitude of small-angle   scattering of charged  particle 
(charge $Z_1$) by  a linear chain of $N$ identical atoms in  the eikonal 
approximation  reads \cite{GLAUBER}
\bea
F_{fi}(q)={ip\over 2\pi}\int d^2{\bf b}\exp(i{\bf qb})
\langle \Psi_f(\left\{\bf r_j\right\})
|1-S({\bf b,s_1,...s_A})|
\Psi_i(\left\{\bf r_j\right\})\rangle,
\label{eq:EL1} 
\eea
where $\Psi_i$ and $\Psi_f$ are the initial and final state wave functions of
the crystal and ${\bf q}$ is the $2D$-vector of the momentum transfer.
The incident particle momentum $p$ is assumed  to be large enough to satisfy
the condition of applicability of the straight paths approximation,
$p/q^2\gg aN$. The latter condition insures the coherence of 
interactions with different atoms. 

The elastic scattering corresponds to $i=f$ and 
 the brackets $\langle\,\,\,\rangle$ signify that an 
average is to be taken over all configurations of atoms in 
the ground state, 
\bea
\langle \Psi(\left\{\bf r_j\right\})|1-S({\bf b,s_1,...s_N})|
\Psi(\left\{\bf r_j\right\})\rangle \nonumber\\
= \int d^3{\bf r_1}...d^3{\bf r_N}|\Psi(\left\{\bf r_j\right\})|^2
\left[1-\exp(i\sum_1^N\chi(\mu|{\bf b-s_j}|))\right].
\label{eq:EL2} 
\eea
In (\ref{eq:EL2}) the total scattering phase is the sum of the phase shifts
contributed by the individual atoms. 
The positions of the $N$ atoms which make up the target 
are defined by the $3D$-vectors ${\bf r_j}$, $j=1,...,N$.
 The $2D$-vectors ${\bf s_j}$ 
are the projections of these vectors on the impact parameter plane.
We neglect all position correlations of the atoms 
and describe the ground state of crystal
by the wave function $|\Psi\rangle$
such that 
\beq
|\Psi(\left\{{\bf r_j}\right\})|^2
= 
\prod_{j=1}^N |\psi({\bf u}_j)|^2\,, \label{eq:Psi}
\endeq
where the $3D$-vectors  ${\bf u}_j$  are defined by 
 ${\bf r}_j=(j-1){\bf a}+{\bf u}_j$, $j=1,...,N$,
 ${\bf a}=(0,\,0,\,a)$ and ${\bf u}_j=({\bf s}_j,z_j)$.

From eq.(\ref{eq:EL2}) it follows that
\bea
F_{ii}(q)=F(q)
={ip}\int bdb J_0(qb)\left\{1-\langle \exp[i\chi({\mu b})]\rangle^N\right\}.
\label{eq:EL5} 
\eea
Hereafter,  $J_{0,1}(x)$ and $K_{0,1}(x)$ are the  Bessel functions and  
the screened Coulomb phase shift function is  
\beq
\chi(\mu b)  =-\beta K_0(\mu b),
 \label{eq:CHI} 
\endeq
with $\beta=2\alpha Z_1Z_2$ and $\mu=r_0^{-1}$.
After integration over longitudinal variables $\{z_j\}$ 
followed by the azimuthal integration the term
$\langle\exp(i\chi)\rangle$ takes the form
\bea
\langle\exp(i\chi)\rangle
=\int d^2{\bf s}\rho({s})
\exp[i\chi(\mu |{\bf b}-{\bf s}|)]\nonumber\\
=\exp(-{\Omega^2b^2})\int_{0}^{\infty} dx
\exp(-x)\nonumber\\ 
\times I_0({2b\Omega\sqrt{x}})\exp[-i\beta K_0(\mu \sqrt{x}/\Omega)]. 
\label{eq:EL8} 
\eea
The $2D$-vector ${\bf s}$, describes the
position of the target atom 
in the impact parameter plane.   
The one-particle probability distribution  $\rho({s})$   is as follows
\beq
\rho({s})=\int dz |\psi({\bf s},z)|^2=
(\Omega^2/\pi)\exp\left(-\Omega^2{\bf s}^2\right). 
\label{eq:RHO}
\endeq
For the most commonly studied elements at room temperature the ratio
$\mu/\Omega$  varies in a wide  range, from $\mu/\Omega\sim 0.1$  
to  $\mu/\Omega\sim 1$ \cite{GEM}. 
Consider first the region of  small impact parameters including 
$b\lsim 1/\Omega$.{\footnote { Notice that  the smallness of the ratio 
${r_A^2/ u^2}\sim 10^{-5}-10^{-6}$,
where $r_A$ is the nuclear radius and $u=1/\Omega$ is the amplitude
of  lattice thermal vibrations,
allows one to neglect the nuclear interactions of the  projectile up to  
$N\sim 10^5$. As we shall see, the absorption effect which we are 
interested in enters the game at much smaller $N$. }} 
For  $ b\lsim 1/2\Omega$ only small $s$,
such that $\mu s\lsim 1$,  contribute.   One can put then
 $K_0(\mu s)\simeq \log(1/\mu s)$
and integrate over $s$,
\bea
\langle \exp(i\chi)\rangle\simeq
\left({\mu\over \Omega}\right)^{i\beta}\exp(-\Omega^2b^2)\nonumber\\
\times\int_{0}^{\infty} dx x^{i\beta/2}\exp(-x)
I_0(2b\Omega\sqrt{x})\nonumber\\
=\left({\mu\over\Omega}\right)^{i\beta}
\Gamma\left(1+{i\beta\over 2}\right)
\!\Phi \left(-{i\beta\over 2};1;-\Omega^2b^2\right)
\label{eq:HB6} 
\eea
In eq.(\ref{eq:HB6})
\bea 
\Phi(a,b;z)=1+{a\over b}{z\over 1!}+{a(a+1)\over b(b+1)}{z^2\over 2!}+...
\label{eq:PHI} 
\eea
is the confluent hyper-geometric function and
$\Phi(a;b;z)=\exp(z)\Phi(b-a;b;-z)$.

From eq.(\ref{eq:HB6}) it follows that 
\bea
\left|\langle \exp(i\chi) \rangle\right|_{b=0}=
\left[{\pi \beta\over 2\sinh(\pi\beta/2)}\right]^{1/2},
\label{eq:EL10} 
\eea
where the identity
\bea
|\Gamma({i\beta/2})|^2={2\pi\over \beta \sinh(\pi\beta/2)}
\label{eq:EL9} 
\eea
has been used.  In the weak coupling regime,  $\beta\ll 1$, 
$$\left|\langle \exp(i\chi) \rangle\right|_{b=0}
\simeq 1- {1\over 2}(\langle\chi^2\rangle-\langle\chi\rangle^2)$$
and 
\bea
\langle\chi^2\rangle-\langle\chi\rangle^2={\pi^2\beta^2\over 24}
\label{eq:HB90}
\eea
For  $\beta\gsim 1$
\bea
\left|\langle \exp(i\chi) \rangle\right|_{b=0}
\simeq \sqrt{\pi\beta}\exp(-\pi\beta /4).
\label{eq:HB9}
\eea
Therefore, at small impact parameters, $b\lsim 1/2\Omega$,
the intensity of outgoing nuclear waves as a function of $N$
 exhibits the exponential attenuation.
In terms of the unitarity cuts of the elastic scattering amplitude
 the  imaginary part of 
the  phase shift function comes from the  cuts 
through the multi-photon projectile-atom blocks
as shown in Fig. 1a. The account of 
diagrams like that of  Fig. 1b which 
allows  cuts only between  projectile-atom blocks, 
gives the scattering matrix of  the form
$\exp(iN\langle \chi\rangle)$ 
and   affects
only the overall real phase of the  amplitude.

\begin{figure}[!htb]
\centering
\epsfig{file=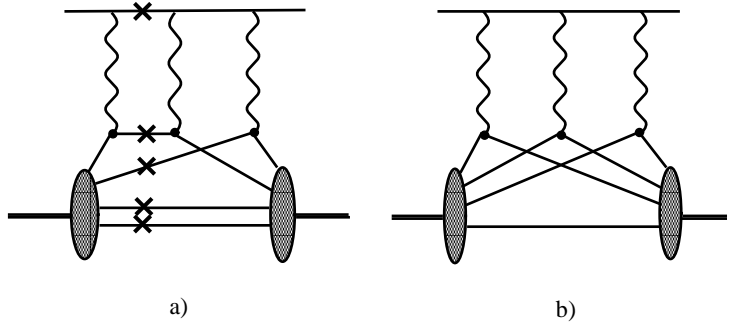,width=12cm}
\vspace{-2.5cm}
\caption{\em   Example of the relevant multiple scattering
diagrams to order $\beta^3$. The unitarity cut of the elastic amplitude $1a)$
 which contributes to the absorption  is shown by crosses. 
The diagram $1b)$ allows cuts only between the projectile-atom blocks
 and does not contribute to 
the absorption effect.}
\label{CUT}
\end{figure}

 The absorption effect becomes weaker toward the  
region of large impact parameters $b\gsim 1/2\Omega$,
\bea
\left|\langle \exp(i\chi) \rangle\right|^N\simeq 
\left|\langle \exp(i\chi) \rangle\right|^N_{b=0}
\left[1+{N\beta^2\over 16}(\Omega b)^4+...\right].
\label{eq:HB10}
\eea  
For still larger $b$,  $b\gg 1/2\Omega$, making use of the
asymptotic form $I_{0}(z)\simeq (2\pi z)^{-1/2}\exp(z)$ yields
\bea
\langle\exp(i\chi)\rangle
\simeq 2\Omega\int {sds\over \sqrt{\pi bs}}\exp[-\Omega^2(b-s)^2]
 \exp[i\chi(\mu s)].
\label{eq:HB2}
\eea
To evaluate the integral (\ref{eq:HB2}) expand $\chi(\mu s)$ in powers 
of $(s-b)$, 
$$\chi(\mu s)\simeq \chi(\mu b)+\omega(s-b).$$
If the frequency $\omega$,
$$
\omega={d\chi\over db}=\mu\beta K_1(\mu b),
$$
is small compared to $\Omega$,
\bea
\omega\ll \Omega,
\label{eq:omega}
\eea 
$$\langle\chi^2\rangle-\langle\chi\rangle^2={\omega^2\over 2\Omega^2}$$
and 
\bea
\langle\exp(i\chi)\rangle\simeq
\exp\left[i\chi-i\omega b\right]\nonumber\\
\times \Omega\int_0^\infty
{sds\over \sqrt{\pi bs}}\exp[-\Omega^2(b-s)^2]\exp[i\omega s]\nonumber\\
\simeq \exp(i\chi)\exp[-\omega^2/4\Omega^2].
\label{eq:HB40} 
\eea
The condition (\ref{eq:omega}) is satisfied 
if $b\gg \beta/\Omega$. For the impact parameters from the region
$\beta/\Omega\ll b \ll 1/\mu$ we may write  $\omega\simeq{\beta/ b}$
and for  larger $b$ such that  $b\gg \mu^{-1}$,
\bea
\omega\simeq
\mu\beta\sqrt{\pi\over 2\mu b}\exp(-\mu b).
\label{eq:omega2}
\eea  
From the consideration presented above it follows that the 
absorption effect in the elastic scattering amplitude
 is especially strong  for  impact parameters
\bea
b\lsim b_{a}= {1\over 2\mu}\log{\pi\mu^2\beta^2 N \over 4\Omega^2}.
\label{eq:RBD} 
\eea
For  $b\ll b_a$ the atomic chain  acts like an opaque ``black'' disc.
Certainly, the value of this finding  differs  for different observables 
and for different  processes proceeding at different impact parameters.
The only thing which is worth noticing here is the representation
of the scattering matrix in the form  
\bea
\langle S(b)\rangle\simeq\exp{\left(iN\chi
-{N\omega^2\over 4\Omega^2}\right)}; \,\,\, b\gg \beta \Omega^{-1}.
\label{eq:SM} 
\eea
The eq.(\ref{eq:SM}) supplemented with the observation that
\bea
\langle S(b)\rangle
\simeq ({\pi\beta})^{N/2}\exp\left(-{N\pi\beta \over 4}\right); 
\,\,\, b\lsim \Omega^{-1}
\label{eq:HB999}
\eea
simplifies all further calculations greatly.

\section{The  elastic cross section.}

Integrating once  by parts reduces $F(q)$ to the form convenient
for evaluation of the total cross section,
\bea
F(q)={ip\mu N\over q}\int_0^{\infty} bdbJ_1(qb)
\langle i\chi^{\prime}\exp(i\chi)\rangle
\langle\exp(i\chi)\rangle^{(N-1)}.
\label{eq:HB8} 
\eea
At small impact parameters, $b\ll 1/2\mu$, 
\bea
\langle\chi^{\prime}\exp(i\chi)\rangle\simeq
\beta\left({\mu\over\Omega}\right)^{i\beta-1}\exp(-\Omega^2b^2)\nonumber\\
\times\Gamma\left({i\beta+1\over 2}\right)
\Phi\left({i\beta+1\over 2};1;\Omega^2b^2\right).
\label{eq:HB91} 
\eea
Because of multiple scatterings  only large impact parameters
 $b$ may contribute 
to $F(q)$ at large $N$ and small $q$.
Hence,
\bea
F(q)
\simeq 
{ip\mu N\over q}\int_{1/\mu}^{\infty} bdbJ_1(qb)
\left[i\chi^{\prime}-\omega\omega^{\prime}/2\Omega^2\right]\nonumber\\
\times\exp(iN\chi)
\exp(-N\omega^2/4\Omega^2) .
\label{eq:EL6} 
\eea
where the explicite form of $\langle\exp(i\chi)\rangle$ at large $b$, 
eq.(\ref{eq:HB40}), has been used. 
 For large $b$, $\omega^2\propto \exp(-2\mu b)$
and  diminishes with growing $b$ 
much faster 
 than  the phase shift function $\chi(\mu b)$ which is 
$\propto \exp(-\mu b)$. One can  see 
that the leading  contribution to the   elastic scattering amplitude
 (\ref{eq:EL6}) comes from 
$$b\sim \mu^{-1}\xi\gg b_{a},$$
where 
$$\xi =\log(\beta N)$$
 and for large $N$ the  
second term in square
brackets in (\ref{eq:EL6}) is small compared to  
the first one.
Then, for  
 $$q\lsim q_0= \mu/\xi$$
 and $\xi\gg 1$, the steepest descent
from the saddle-point
\bea
b_0=\mu^{-1}[\xi+i\pi/2]
\label{eq:B}
\eea
in eq.(\ref{eq:EL6}) yields
\bea
F(q)\simeq {ipb_0\over q}J_1(qb_0).
\label{eq:FQ0}
\eea
The effect of lattice thermal vibrations at small $q$ 
appears to be  marginal and reduces to the factor
$\exp({\mu^2/ 4\Omega^2 N})$ in (\ref{eq:FQ0})
which is irrelevant at large $N$. The amplitude $F(q)$ at $q\to 0$  
coincides with the elastic scattering amplitude given by the SL 
approximation \cite{KALASH}.

If  $q\gsim q_0$ the stationary phase approximation gives the elastic
 scattering amplitude of the form
\bea
F(q)\simeq {-ip\sqrt{\eta}\over \mu q }
\exp\left(-{iq\eta\over \mu}\right)
\exp\left(-{q^2\over 4\Omega^2 N}\right)
\label{eq:FQQ}
\eea
 where $\eta=\log(\mu\beta N/q) \gg 1$. 
The account  of the lattice thermal vibrations  
insures the convergence of the integral  for  the  coherent elastic scattering 
cross section,
\bea
\sigma_{el}={\pi\over p^2} \int {dq^2}|F(q)|^2
\approx {\pi \xi^2\over \mu^2}\int_0^{q_0^2}{dq^2\over q^2}
J^2_1\left({q\xi\over\mu}\right)\nonumber\\
+ {\pi\over \mu^2}\int_{q_0^2}^{\infty}
{dq^2\over q^2}\log\left({\mu\beta N\over q}\right)\exp
\left(-{q^2\over 2\Omega^2 N}\right),
\label{eq:SIGEL}
\eea 
which for $\xi\gg 1$ is simply
\bea
\sigma_{el}
\approx {\pi \over \mu^2}\xi^2.
\label{eq:SIGEL1}
\eea 

\section{The quasi-elastic cross section.}
In this paper we focus on the coherent nucleus-atom interactions.
The incoherent process of  ionization of the target atom   is
suppressed by the factor  $\sim Z^{-1}_2$. Then,
the inelastic process  which via unitarity gives rise to the 
attenuation of  elastic amplitude 
is the process of the    quasi-elastic 
scattering (Fig.1a). Its cross section is the quantity \cite{GLAUBER}
\bea
p^2{{d\sigma_{Qel}}\over {d^2{\bf q}}}=\sum_{f}|F_{fi}(q)|^2-
|F_{ii}(q)|^2
\label{eq:QEL1}  
\eea
where the sum extends over all final states of crystal, in which no
particle production takes place. 
The closure  relation then
yields
\bea
{{d\sigma_{Qel}}\over {d^2{\bf q}}}=
{1\over 4\pi^2}\int d^2{\bf b}d^2{\bf b}^{\prime}
\exp[{i{\bf q}({\bf b}-{\bf b}^{\prime})}]\nonumber\\
\times\left\{\langle \exp[{i\chi(\mu b)-
i\chi^*(\mu b^{\prime})}]\rangle^N\right.\nonumber\\
\left.-\langle \exp[{i\chi(\mu b)}]\rangle^N
\langle \exp[{-i\chi^*(\mu b^{\prime})}]\rangle^N\right\}
\label{eq:QEL2}  
\eea
and
\bea
\sigma_{Qel}=\int d^2{\bf b}\left\{1-
\left|\langle\exp[{i\chi(\mu b)}]\rangle\right|^{2N}\right\}.
\label{eq:QEL3}  
\eea
In the SL approximation $\left|\langle\exp[{i\chi(\mu b)}]\rangle\right|=1$
and $\sigma_{Qel}=0$.
From (\ref{eq:SM}) and the  discussion
of the absorption radius, $b_a$, presented above 
 it follows that for  $\xi\gg 1$ 
\bea
1-
\left|\langle\exp[{i\chi(\mu b)}]\rangle\right|^{2N}\approx \theta(2b_a-b)
\label{}
\eea
and  
\bea
\sigma_{Qel}
\approx \pi (2b_a)^2
\approx {\pi\over \mu^2} \xi^2
\label{eq:QEL4}  
\eea

\section{The total cross section.}

From eq.(\ref{eq:FQ0}) by means of the optical theorem we find 
the total cross section 
\bea
\sigma_{tot}={4\pi\over p}Im F(0)
\approx {2\pi\over\mu^{2}} \xi^2. 
\label{eq:STOT}
 \eea
Thus, we conclude that at high energy and  in the large-$N$ limit
\bea
\sigma_{el}\approx \sigma_{Qel}\approx{1\over 2}\sigma_{tot}. 
\label{eq:SELTOT}
\eea

\section{Coulomb excitation of ultra-relativistic particles
and nuclei in crystal channel. 
The excitation cross section to the lowest order. The Born approximation.}

Consider now the process of the 
 coherent Coulomb excitation of ultra-relativistic
particles and  nuclei passing through the  crystal.
This way of the experimental study of rare processes  has been proposed 
in \cite{PROP1, PROP2,PIVOVAR,PROP3,PROP4,FUSINA,DUBIN}.

 The  ultra-relativistic
projectile-nucleus 
(the mass number $A$, the  charge $Z_1$ and  the four-momentum $p$) 
 moving along a crystal  axis undergoes a correlated series of 
 soft  collisions
which give rise to diagonal ($A\to A$, $A^*\to A^*$) and
 off-diagonal ($A\to A^*$, $A^*\to A$)
 transitions. 
 
 In \cite{FUSINA,PROP1,PROP2} it has been proposed to study
the electric dipole transition in $^{19}F$,
  the excitation of the state $|J^{\pi}={1/2}^-\rangle$ 
from the ground state $|1/2^+\rangle$. 
The phenomenological   matrix element of the transition $1/2^+\to 1/2^-$
is \cite{NUCSTAT}
\beq
{\cal M}={1\over 2}d\bar u(p^{\prime}) 
\gamma_5 \left(\hat q\hat\varepsilon- 
\hat\varepsilon\hat q\right) u(p), 
\label{eq:M}
\endeq
where both $u(p^{\prime})$ and $ u(p)$ are bispinors of initial
 and final states of the
projectile,
 $d$ is the transition
 dipole moment and  $\varepsilon$
 is the photon polarization vector. 
The  transverse and longitudinal components the 4-vector 
$p-p^{\prime}$ are  denoted by  ${\bf q}$ and $\kappa$, respectively.
In what follows $q=|{\bf q}|$.The only phenomenological parameter of 
the problem is the dipole moment $d$.  
 The measured life-time  of the $110$ KeV   level $^{19}F(1/2^-)$ is
$\tau=(0.853\pm 0.010)\times 10^{-9}$ sec \cite{AJZEN} and
 the dipole moment of the $1/2^+\to 1/2^-$ transition,
determined from the width  of the  level $^{19}F(1/2^-)$
 is  $d\simeq 5\times 10^{-8}$ KeV$^{-1}$ \cite{NUCSTAT}. 
Then, first,
because of large value of $\tau$
the decay of excited state inside the target crystal can be safely neglected 
and, 
second, 
due to the smallness of $d$, the excitation amplitude
is much smaller than the elastic Coulomb 
amplitude for all $q$ up to $q\sim \sqrt{4\pi\alpha}Z_1/d$ and can
 be considered as a perturbation. Thus, the multi-channel problem
reduces to the one-channel one.

The high-energy helicity-flip Born amplitude 
 of the transition $1/2^+\to 1/2^-$ in  collision of the 
projectile-nucleus with $N$  bound atoms in crystal reads 
\beq
{F}^{B}_{ex}({\bf q})=S(\kappa){p\over 2\pi}
{g(\bsigma{\bf q})
\over q^2 +\lambda^2 }\exp\left(-{q^2\over 4\Omega^2}\right)\,,               
\label{eq:TB}
\endeq 
where  $\bsigma=(\sigma_1,\sigma_2,\sigma_3)$ 
is the Pauli spin vector, 
$\{\sigma_i,\sigma_j\}=2\delta_{ij}$ and the amplitude we are constructed
 is to be regarded as an operator 
which transforms the initial helicity state of the projectile into its final 
state. 
In the denominator of eq.(\ref{eq:TB})
 $\lambda^2=\mu^2+\kappa^2$.
In the Glauber approximation the longitudinal momentum transfer
which determines the coherency length, $l_c\sim \kappa^{-1}$,
 reads  \cite{GRIBOV}
\beq
\kappa={M\Delta E\over p},
\label{eq:kappa}
\eeq
where $M$ is  the mass of  
 projectile and   $\Delta E$ is the excitation energy 
\footnote{The Fresnel corrections
to the eikonal approximation which are neglected here  become important
at large $N$ or at large $q$ diminish the coherency length 
and  bring about an additional suppression of coherent processes
\cite{NEUT}.}.

 The  structure factor of crystal, 
$S(\kappa)$, to the first order in $g$   is  
\beq
S(\kappa)=\exp\left[-{\kappa^2\over 4\Omega^2} \right]
{{\sin(\kappa Na/2)}\over{\sin(\kappa a/2)}}\,.
 \label{eq:SLNEW}
\endeq 
If the projectile momentum  satisfies 
 the resonance condition \cite{PROP1,PROP2,FUSINA,PROP3} 
\beq
 {M\Delta E\over p}={2\pi n\over a}\,,\,\,\,n=0,\,1,\, 2...\label{eq:KAPRES}
\endeq
$S(\kappa)\sim N$.
 Then, to the first order in $g$ and to the zero order in $\beta$
(Born approximation) the cross section of the coherent
excitation of the projectile in scattering on a chain of $N$ atoms in 
crystal is     
\bea 
\sigma^B_{ex}={\pi\over p^2}\int dq^2 |F^B_{ex}({\bf q})|^2\nonumber\\
\sim {g^2N^2\over 4\pi}\left[
\log\left(1+{2\Omega^2\over\lambda^2}\right)-
{2\Omega^2\over\lambda^2+2\Omega^2}\right],
\label{eq:SIGBORN}
\eea
where 
$g=\sqrt{4\pi\alpha}dZ_2$.
The central idea of \cite{PROP1, PROP2,PIVOVAR,PROP3,FUSINA,DUBIN} 
based on the Born approximation is that the  transition rate can be 
enhanced substantially due to coherency
of interactions which is assumed  to sustain  over the large distance scale. 
 The law  $\sigma_{ex}\propto\,N^2$
is expected to hold true  up to the crystal thicknesses    
$N=L/a\sim 10^5-10^6$ in tungsten target.
In \cite{DUBIN} the Born approximation for the coherent excitation
of $\Sigma^+$ in high-energy  proton-crystal interactions
$p\gamma\to \Sigma^+ $  has been assumed 
to be valid up to $N\sim 10^8$. However, the account of the initial and final
state  Coulomb interactions  dramatically 
changes the dependence of $\sigma_{ex}$ on $N$. For instance, 
at $N=2$  the excitation amplitude is of the 
following form
\beq
F^{(2)}_{ex}({\bf q })={p\over \pi}
\int {d^2{\bf b}}\exp(i{\bf q} {\bf b})
\langle {f}^B_{ex}
\exp(i\chi)\rangle
\langle \exp(i\chi)\rangle
\label{eq:TFULL2} 
\endeq
The first one  of two bracketed factors in eq.(\ref{eq:TFULL2})
 corresponds to the nuclear excitation amplitude
in scattering on  a bound atom.
It differs 
from  the excitation amplitude of the  Born approximation, $f_{ex}^B({\bf b})$,
by the  multiplicative phase factor which 
 is due 
to the initial and final state multiple Coulomb scattering. 
At small impact parameters, $b\lsim 1/2\Omega $,
\bea
\langle {f}_{ex}^{B} \exp(i\chi)\rangle\simeq
S(\kappa){g\over 2\pi b}{(\bsigma {\bf n_b})}
\sinh\left({1\over 2}\Omega^2b^2\right)\exp\left(-{1\over 2}\Omega^2b^2\right).
\label{eq:GB22}
\eea
For large $b$
\bea
\langle {f}_{ex}^B \exp(i\chi)\rangle=
S(\kappa){g\over 4\pi}\int d^2{\bf s}\rho(s)
(\bsigma ({\bf n_b}-{\bf n_s}))\nonumber\\
\times\lambda K_1(\lambda |{\bf n_b}-{\bf n_s}|   )
\exp\left[i\chi\left(\mu |{\bf n_b}-{\bf n_s}|\right) \right]\nonumber\\
\simeq
S(\kappa){g\over 4\pi}(\bsigma {\bf n_b})\lambda K_1(\lambda b)
\exp(i\chi)\exp\left(-{\omega^2\over 4\Omega^2}\right),
\label{eq:GB2}
\eea
where ${\bf n_b}={\bf b}/|{\bf b}| $, ${\bf n_s}={\bf s}/|{\bf s}|$
and $b\gsim \mu^{-1}$. As far as for small $b$
$$|\langle {f}_{ex}^B \exp(i\chi)\rangle|^2 \propto
\Omega^2b^2$$
and for large impact parameters, $b\gsim 1/\mu$,
$$bK_1^2(\mu b)\propto \exp(-2\mu b)$$
the cross section
\bea
\sigma^{(2)}_{ex}=\int d^2{\bf b}
|\langle {f}_{ex}^B \exp(i\chi)\rangle|^2
|\langle \exp(i\chi)\rangle|^2 \nonumber\\
\simeq 4\sigma^{(1)}_{ex}\left(1-{\omega^2\over 2\Omega^2}\right)  
\label{eq:SEC2}
\eea
is dominated by $b\sim 1/2\mu$. For the diamond crystal
$\mu/\Omega\simeq 0.16$ \cite{GEM}.
Hence, $\omega^2/2\Omega^2\simeq 2\beta^2\mu^2/\Omega^2\sim 1/20$.
This estimate shows that even  for the diamond crystal target  
the Born approximation is irrelevant  already  at $N\gsim  10$.

\section{The  multiple scattering effects and absorption in
the coherent Coulomb excitation processes}

 The   transition amplitude on a chain 
 of $N$ identical atoms  including all the multi-photon
 t-channel exchanges reads  
\beq
F_{ex}({\bf q })={p\over \pi}
\int {d^2{\bf b}}\exp(i{\bf q} {\bf b})
\langle {f}^B_{ex}
\exp(i\chi)\rangle
\langle \exp(i\chi)\rangle^{N-1}
\label{eq:TFULL} 
\endeq
 Because of both the multiple scattering effect and absorption   only 
 large impact parameters, $b\gg \mu^{-1}$,
 may contribute to $F_{ex}({\bf q })$.
Then,the evaluation of $F_{ex}(q)$ reads
\bea
F_{ex}({\bf q })\approx {gp\over 2\pi}S(\kappa)
(\bsigma{\bf n_q})\int_{1/\mu}^{\infty} 
bdbJ_1(qb)\nonumber\\
\times
\lambda K_1(\lambda b)\exp(iN\chi)\exp(-N\omega^2/4\Omega^2), 
\label{eq:TREDUCE} 
\eea
where  ${\bf n_q}={\bf q}/|{\bf q}|$. 
The contribution of the domain
$q\lsim q_0=\mu/\xi$  to 
the excitation  cross section 
can be neglected as far as  $F_{ex}\propto q$ in this region. 
If $q\gg q_0$ and $\xi\gg 1$,
 the stationary phase approximation gives the coherent excitation
 amplitude of the form
\bea
F_{ex}({\bf q })\approx 
{ipg(\bsigma{\bf n_q})\over 2\pi \beta}{S(\kappa)\over N}
{\lambda\over\mu}{\sqrt{\eta}}
\exp\left(-\delta\eta\right)\nonumber\\
\times\exp\left(-{iq\eta\over \mu}\right)
\exp\left(-{q^2\over 4\Omega^2 N}\right).
\label{eq:TLARGE}
\eea
We see that the helicity-flip dynamics removes the factor $1/q$
from the elastic  amplitude (\ref{eq:FQQ}) thus making the 
UV-regularization
of the excitation cross section indispensable.
The latter is evaluated as,
\bea
\sigma_{ex}={\pi\over p^2}\int dq^2|F_{ex}({\bf q })|^2\nonumber\\ 
\sim  {g^2N^{1-\delta}\over 8\pi}C
\log\left({N\over\delta\gamma}\right),
\label{eq:SIGEX}
\eea
where $C=\gamma^{\Delta}\Delta^2\Gamma(\Delta)$, 
$\gamma=2\Omega^2/\beta^2\mu^2$, $\Delta=\lambda/\mu$ 
and $\delta =\Delta - 1\sim \kappa^2/2\mu^2\ll 1$. In (\ref{eq:SIGEX})
we put simply $S(\kappa)=N$.
Thus, the account of  multiple scatterings and absorption turns
the Born approximation cross section   
$\sigma_{ex}\propto N^2$ into $\sigma_{ex}\propto N^{1-\delta}\log N$.
 In the limit of $p\to \infty$ and $\delta\to 0$,
\bea
\sigma_{ex}
\sim  {g^2N\over 8\pi}\gamma
\log\left({N\over \gamma}\right)
\label{eq:SIGEX0}
\eea
The dependence of $\sigma_{ex}$ on $N$ differs from that of the fully
unitarized elastic cross section,
 $\sigma_{el}\propto \log^2 N$. The reason is that
  in $\sigma_{ex}$ we sum 
the eikonal diagrams to all orders in $\beta$ but only to the first order
in $g$. 
Such a  procedure of unitarization  is, of course, incomplete, but 
 this is of no importance for practical purposes since
the smallness of $d^2\Omega^2$ makes the next to leading order terms 
negligibly small up to $N\sim \alpha Z_1^2/\delta\Omega^2d^2\sim 10^{12}$.

\section{Summary}
The main goal we pursued in this paper is a consistent description
of the coherent Coulomb excitation of ultra-relativistic particles
and nuclei passing through the aligned crystal. We started with the discussion
of the elastic scattering and found that the account of the lattice thermal
vibrations within the Glauber multiple scattering theory gives rise
to the strong absorption effect.The radius of the absorption region
in the impact parameter space appeared to grow logarithmically with 
growing crystal thickness.
We derive convenient representation for the 
scattering matrix with absorption and calculate the coherent elastic 
and  the incoherent quasi-elastic cross sections.
  The suppression of scattering amplitudes in the absorption
region is shown to  serve as a natural UV regulator and enables 
consistent calculation
of the  cross section of the coherent nuclear excitation, $\sigma_{ex}$.
The dependence of $\sigma_{ex}$ on the crystal thickness is found.
The multiple scattering effects are shown to become numerically important
already at $N\gsim 1$ thus leaving no room  for 
the Born approximation widely used in early analyses of the problem.

{\bf Acknowledgments:}   Thanks are due to N.N. Nikolaev for
useful comments.


\begin{thebibliography}{99}

\bibitem{GLAUBER}
R.J. Glauber, in Lectures in Theoretical Physics, edited by W.E. Brittin
et al., Interscience Publishers, Inc., New York, vol.1, p. 315, 1959.

\bibitem{KALASH}
N.P.Kalashnikov, E.A.Koptelov and M.I.Ryazanov, 
Sov. Phys. JETP 63 (1972) 1107;\\
N.P.Kalashnikov and V.D.Mur, 
Sov. J. Nucl. Phys. 16 (1973) 613.

\bibitem{GEM}
D.S. Gemmel, Rev.  Mod. Phys. 46 (1974) 1.



\bibitem{PROP1}
V.V. Okorokov, Sov. J. Nucl. Phys. {\bf 2} (1966) 719;
 V.V. Okorokov, Yu.L. Pivovarov, A.A. Shirokov, S.A. Vorobev,
{\sl Proposal of experiment on coherent excitation of relativistic nuclei in
 crystals},
Moscow,  ITEP-90-49, Fermilab Library Only.

\bibitem{PROP2}
V.V. Okorokov, S.V. Proshin,
{\sl Investigation of the coherent excitation of the relativistic nuclei in 
a crystal}, Moscow, ITEP-13-1980 

\bibitem{PIVOVAR}
Yu.L. Pivovarov, H. Geissel,
 Yu.M. Filimonov, O.E. Krivosheev, C.Scheidenberger,
{\sl  On the resonant coherent excitation of relativistic heavy ions}
  GSI-95-38, Darmstadt, Jul 1995 

\bibitem{PROP3}
Yu.L. Pivovarov and A.A. Shirokov, Sov. J. Nucl. Phys. {\bf 37} (1983) 653

\bibitem{PROP4}
Yu.L. Pivovarov, A.A. Shirokov, S.A. Vorobev,  
Nucl.Phys. {\bf A509} (1990) 800  

\bibitem{FUSINA}
R. Fusina, J.C. Kimball,
Nucl.Instrum.Meth. {\bf B33} (1988) 77 

 
\bibitem{DUBIN}
A. Dubin, 
Sov.J.Nucl.Phys.52:790-793,1990, Yad.Fiz.52:1243-1249,1990 


\bibitem{GRIBOV}
V.N. Gribov, Sov. Phys. JETP 29 (1969) 483; 30 (1970) 709.

\bibitem{NEUT}
V.R. Zoller, Phys.Lett. {\bf B416} (1998) 447

\bibitem{AJZEN}
F. Ajzenberg-Selove, Nucl. Phys A190 (1972) 1.

\bibitem{NUCSTAT}
V.R. Zoller, JETP Lett. {\bf 75 } (2002) 119 ; Pis'ma v ZhETP 
{\bf 75 } (2002) 147.




\end{thebibliography}
\end{document}